\documentclass[twocolumn,showpacs,prc,floatfix]{revtex4}
\usepackage{graphicx}
\usepackage[dvips,usenames]{color}
\usepackage{amsmath}
\def\beq{\begin{eqnarray}} \def\eeq{\end{eqnarray}}
\def\beqstar{\begin{eqnarray*}} \def\eeqstar{\end{eqnarray*}}
\newcommand{\bal}{\begin{align}} \def\eal{\end{align}}
\newcommand{\beqe}{\begin{equation}} \newcommand{\eeqe}{\end{equation}}
\newcommand{\p}[1]{(\ref{#1})}

\begin {document}
\title{Spin ordered phase transitions in\\ isospin asymmetric nuclear matter}
\author{ A. A. Isayev}
 \affiliation{Kharkov Institute of
Physics and Technology, Academicheskaya Street 1,
 Kharkov, 61108, Ukraine
 }
 \date{\today}
\begin{abstract}   Spin polarized
states in nuclear matter with Skyrme effective forces are studied
on the base of  a Fermi liquid theory  for a wide range of isospin
asymmetries and densities. There are a few possible scenarios of
spin ordered phase transitions: (a) nuclear matter with  SLy4
interaction undergoes at some critical density a phase transition
to a spin polarized state with the oppositely directed spins of
neutrons and protons; (b) for  SkI5 interaction, a spin polarized
state with the like-directed neutron and proton spins is formed;
(c) nuclear matter with SkI3 interaction under increasing density,
at first, undergoes a phase transition to the state with the
opposite directions of neutron and proton spins, which goes over
at larger density to the state with the same direction of nucleon
spins. Spin polarized states at strong isospin asymmetry are
accompanied by the long tails in the density profiles of the
neutron spin polarization parameter near the critical density, if
the energy gain of the transition from the nonpolarized state to a
polarized one is the decreasing function of isospin asymmetry
(SLy4 force). If the energy gain is increased with isospin
asymmetry, there are no long tails in the density distribution of
the neutron spin polarization parameter (SkI3, SkI5 forces).
\end{abstract}
\pacs{21.65.+f; 26.60.+c; 75.25.+z; 71.10.Ay} \maketitle

 The issue of spontaneous appearance of  spin
polarized states in nuclear matter is a topic of a great current
interest due to its relevance in astrophysics. In particular, the
scenarios of supernova explosion and cooling of neutron stars are
essentially different, depending on whether nuclear matter is spin
polarized or not. On the one hand, the models with the effective
Skyrme and Gogny nucleon-nucleon (NN) interaction predict the
occurrence of spin instability in nuclear matter at densities in
the range from $\varrho_0$ to $6\varrho_0$ for different
parametrizations of the NN potential~\cite{R}--\cite{RPV}
($\varrho_0=0.16\,\mbox{fm}^{-3}$ is the nuclear  saturation
density). On the other hand, for the models with the realistic NN
interaction, the ferromagnetic  phase transition seems to be
suppressed up to densities well above
$\varrho_0$~\cite{PGS}--\cite{KS}.

 Here we continue
the research of spin polarizability of nuclear matter with the use
of an effective NN interaction. The main objective is to study the
possible scenarios of spin ordered phase transitions in nuclear
matter with Skyrme forces, attracting parametrizations of a NN
potential being relevant  for calculations at strong isospin
asymmetry and high density.    In particular, we choose SLy4
effective interaction, constructed originally to reproduce the
results of microscopic neutron matter calculations~\cite{CBH}. We
utilize SkI3 and SkI5 parametrizations as well, giving a correct
description of isotope shifts in neutron-rich medium and heavy
nuclei~\cite{RF}. As compared with the research of Ref.~\cite{IY}
with SLy4 and SLy5 effective interactions, here we explore a wider
domain of isospin asymmetries, including symmetric nuclear matter
and neutron matter as limiting cases. Besides, we use also SkI3
and SkI5 parametrizations, that will allow us to study new
scenarios of spin ordered phase transitions, not found in
Ref.~\cite{IY}.

 The basic formalism is presented in detail
in Ref.~\cite{IY}.  We are interested in studying   spin polarized
states with like-directed and oppositely directed spins of
neutrons and protons. One should solve the self-consistent
equations for the coefficients
$\xi_{00},\xi_{30},\xi_{03},\xi_{33}$ in the expansion of the
single particle energy in Pauli matrices in spin and isospin
spaces \bal\xi_{00}({\bf p})&=\varepsilon_{0}({\bf
p})+\tilde\varepsilon_{00}({\bf p})-\mu_{00},\;
\xi_{30}({\bf p})=\tilde\varepsilon_{30}({\bf p}),\label{14.2} \\
\xi_{03}({\bf p})&=\tilde\varepsilon_{03}({\bf p})-\mu_{03}, \;
\xi_{33}({\bf p})=\tilde\varepsilon_{33}({\bf
p}).\nonumber\end{align} Here $\varepsilon_0({\bf p})$ is the free
single particle spectrum,  and
$\tilde\varepsilon_{00},\tilde\varepsilon_{30},\tilde\varepsilon_{03},\tilde\varepsilon_{33}$
are the Fermi liquid (FL) corrections to the free single particle
spectrum, related to the normal FL amplitudes $U_0({\bf
k}),...,U_3({\bf k}) $ by formulas
\begin{align}\tilde\varepsilon_{00}({\bf
p})&=\frac{1}{2\cal V}\sum_{\bf q}U_0({\bf k})f_{00}({\bf
q}),\;{\bf k}=\frac{{\bf p}-{\bf q}}{2}, \label{14.1}\\
\tilde\varepsilon_{30}({\bf p})&=\frac{1}{2\cal V}\sum_{\bf
q}U_1({\bf k})f_{30}({\bf q}),\nonumber\\ 
\tilde\varepsilon_{03}({\bf p})&=\frac{1}{2\cal V}\sum_{\bf
q}U_2({\bf k})f_{03}({\bf q}), \nonumber\\
\tilde\varepsilon_{33}({\bf p})&=\frac{1}{2\cal V}\sum_{\bf
q}U_3({\bf k})f_{33}({\bf q}). \nonumber
\end{align}

The distribution functions $f_{00},f_{03},f_{30},f_{33}$, in turn,
can be expressed in terms of  the components $\xi$ of the single
particle energy  and satisfy the normalization conditions for the
total density $\varrho_n+\varrho_p=\varrho$, excess of neutrons
over protons $\varrho_n-\varrho_p\equiv\alpha\varrho$,
ferromagnetic (FM)
$\varrho_\uparrow-\varrho_\downarrow\equiv\Delta\varrho_{\uparrow\uparrow}$
and antiferromagnetic (AFM)
$(\varrho_{n\uparrow}+\varrho_{p\downarrow})-
(\varrho_{n\downarrow}+\varrho_{p\uparrow})\equiv\Delta\varrho_{\uparrow\downarrow}$
spin order parameters, respectively  ($\alpha$ being the isospin
asymmetry parameter,
$\varrho_\uparrow=\varrho_{n\uparrow}+\varrho_{p\uparrow}$ and
$\varrho_\downarrow=\varrho_{n\downarrow}+\varrho_{p\downarrow}$,
with  $\varrho_{n\uparrow},\varrho_{n\downarrow}$
 and
 $\varrho_{p\uparrow},\varrho_{p\downarrow}$ being the neutron and
 proton number densities with spin up and spin down).
  The quantities of interest are  the neutron and
proton spin polarization parameters
\begin{equation}
\Pi_n=\frac{\varrho_{n\uparrow}-\varrho_{n\downarrow}}{\varrho_n},\quad
\Pi_p=\frac{\varrho_{p\uparrow}-\varrho_{p\downarrow}}{\varrho_p},\nonumber
\end{equation}
characterizing spin ordering in neutron and proton subsystems. In
 the numerical solution of the
self-consistent equations we utilize SLy4, SkI3 and SkI5 Skyrme
forces. Note that the density dependence of the nuclear symmetry
energy, calculated with these Skyrme interactions, gives the
neutron star models in a broad agreement with the
observables~\cite{SMK}. Another important constraint on the set of
Skyrme force parameters can be obtained, if to consider expression
for the effective mass of a neutron $m_n^{*}$  in  totally spin
polarized neutron matter
\begin{equation}
\frac{m_0}{m_n^{*}}=1+\frac{\varrho
m_0}{\hbar^2}t_2(1+x_2),\label{nem}
\end{equation}
where $m_0$ is the bare mass of a nucleon. Eq.~\p{nem} follows
from expressions for FL amplitudes in neutron matter~\cite{IY}.
Since usually for Skyrme parametrizations $t_2<0$, then we get the
constraint $x_2\leq-1$, which guarantees the stability of totally
polarized neutron matter at high densities~\cite{RPV,KW94}. The
Skyrme parametrizations SLy4, SkI3 and SkI5 satisfy this
condition.

\begin{figure}[tb] 
\begin{center}
\includegraphics[height=13.8cm,width=8.6cm,trim=49mm 86mm 56mm 46mm,
draft=false,clip]{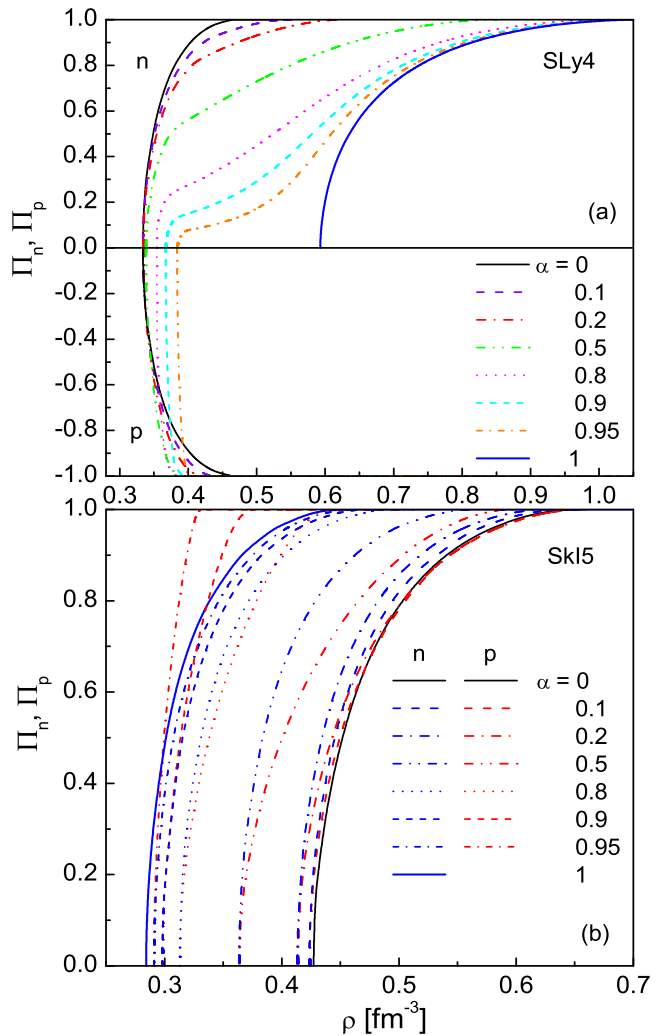}\end{center} \caption{(Color online)
Neutron  and proton  spin polarization parameters as functions of
density  at zero temperature for (a) SLy4 force and (b) SkI5
force. } \label{fig1}
\end{figure}

Fig.~\ref{fig1}a shows the density dependence of the neutron and
proton spin polarization parameters at zero temperature for SLy4
force. The main qualitative feature is that for SLy4 force there
are only solutions corresponding to the oppositely directed spins
of neutrons and protons in a spin polarized state and there are no
solutions corresponding to the same direction of neutron and
proton spins. The reason is that for SLy4 force the FL amplitude
$U_1$, determining spin--spin correlations, becomes more and more
repulsive with  the increase of nuclear matter density, while the
 FL amplitude $U_3$, describing spin--isospin correlations, becomes more and more attractive with
 density, leading to spin instability with the oppositely directed
spins of neutrons and protons at higher densities. The critical
density of spin instability in symmetric nuclear matter
($\alpha=0$), corresponding to AFM spin ordering
($\Delta\varrho_{\uparrow\downarrow}\not=0$,
$\Delta\varrho_{\uparrow\uparrow}=0$), is $\varrho_c\approx0.33$
fm$^{-3}$. It is less than the critical density of FM instability
 in neutron matter, $\varrho_c\approx0.59$ fm$^{-3}$. Even small
admixture of protons to neutron matter leads to the appearance of
long tails in the density profiles of the neutron spin
polarization parameter near the transition point to a spin ordered
state. As a consequence, a spin polarized state is formed much
earlier in density than in pure neutron matter.

As seen from Fig.~\ref{fig1}b, for SkI5 force, oppositely to the
case with SLy4 force,  there are only solutions corresponding to
the  same direction of neutron and proton spins in a polarized
state and  there are no solutions corresponding to their opposite
directions. Comparing to the previous case, the FL amplitudes
$U_1$ and $U_3$ exchange their roles: the FL amplitude $U_3$
becomes more and more repulsive with the increase of nuclear
matter density, while the
 FL amplitude $U_1$ becomes more and more attractive with
 density, leading to spin instability with the like-directed
spins of neutrons and protons at higher densities. For SkI5 force,
a phase transition to the  FM spin  state in neutron matter takes
place at the critical density $\varrho_c\approx0.28$ fm$^{-3}$. It
is less than the critical density of spin instability in symmetric
nuclear matter $\varrho_c\approx0.43$ fm$^{-3}$, corresponding to
 FM spin ordering ($\Delta\varrho_{\uparrow\uparrow}\not=0$,
$\Delta\varrho_{\uparrow\downarrow}=0$). An important peculiarity
is that there are no long tails in the density profiles of the
neutron spin polarization parameter at large isospin asymmetry.
Hence, in this case a small admixture of protons to neutron matter
doesn't considerably change the critical density of spin
instability and even leads  to its increase.

\begin{figure}[tb] 
\begin{center}
\includegraphics[height=12.6cm,width=8.6cm,trim=49mm 102mm 54mm 46mm,
draft=false,clip]{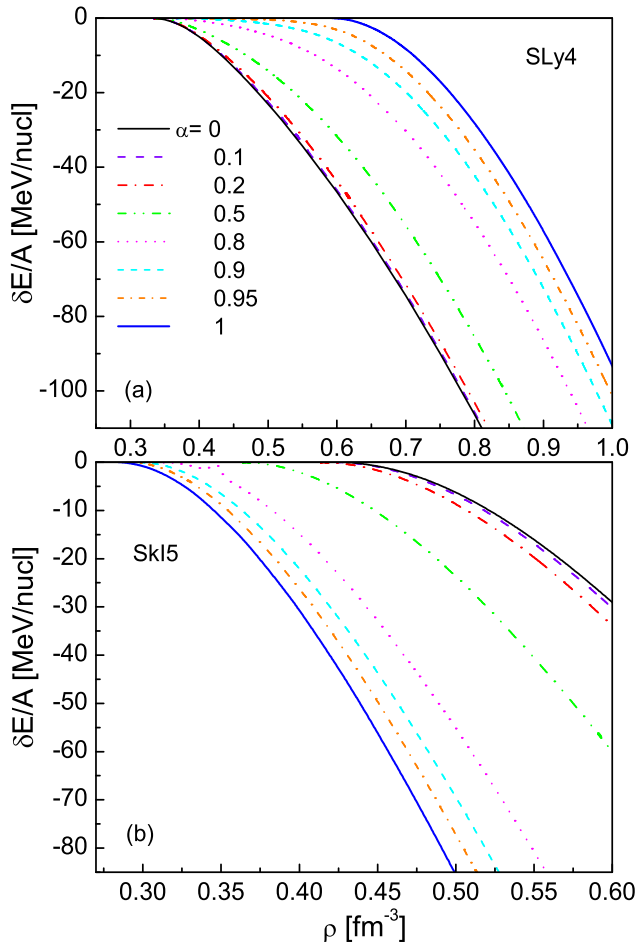}
\end{center}
\caption{(Color online) Total energy   per nucleon, measured from
its value in the normal state,   as a function of density at zero
temperature for (a) SLy4 force and (b) SkI5 force. } \label{fig2}
\end{figure}

 In Fig.~\ref{fig2} the total energies per nucleon  of the spin ordered and
nonpolarized states are compared at zero temperature for SLy4 and
SkI5 forces. One can see that nuclear matter undergoes a phase
transition to the state with the oppositely directed (SLy4 force)
or like-directed (SkI5 force)
 spins of neutrons and protons at
some critical density, depending on  isospin asymmetry. It is
worth to note an important difference in isospin dependences for
these two cases. For SLy4 interaction, the difference between the
total energies of  spin polarized  and nonpolarized  states is
largest at the given density for symmetric nuclear matter while
for SkI5 interaction it is largest for neutron matter. This means
that a phase transition in density  to a spin polarized state will
take place earlier in symmetric nuclear matter than in neutron
matter for SLy4 force, while for SkI5 force, oppositely, it will
occur earlier in neutron matter,  as was mentioned above.

\begin{figure}[tb] 
\begin{center}
\includegraphics[height=12.9cm,width=8.6cm,trim=49mm 95mm 56mm 46mm,
draft=false,clip]{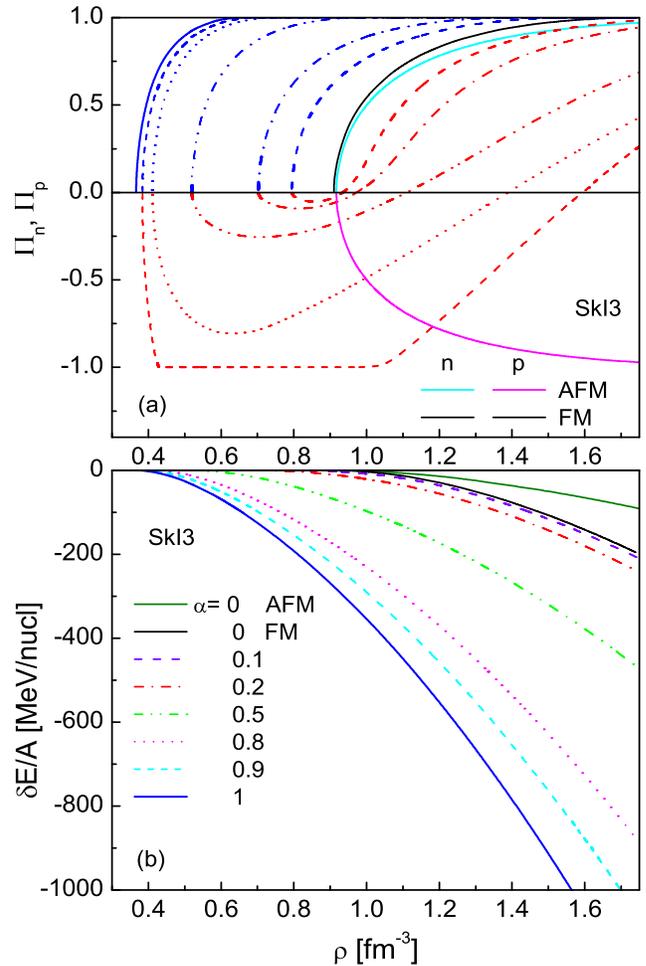}\end{center}
 \caption{(Color online) The dependences for SkI3 force: (a) Same as in
Fig.~\ref{fig1} with the legends of Fig.~\ref{fig1}b; (b) same as
in Fig.~\ref{fig2}. Also the curves, corresponding to FM and AFM
spin ordering in symmetric nuclear matter, are shown.}
\label{fig3}
\end{figure}
Fig.~\ref{fig3}a shows the neutron and proton spin polarization
parameters as functions of density at zero temperature for SkI3
force.  There are two types of solutions of the self-consistent
equations in symmetric nuclear matter, corresponding to FM and AFM
ordering of neutron and proton spins. Due to proximity of FL
amplitudes $U_1$ and $U_3$, the respective critical densities are
very close to each other ($\varrho_c\approx 0.910\,\mbox{fm}^{-3}$
for FM ordering and $\varrho_c\approx 0.917\,\mbox{fm}^{-3}$  for
AFM ordering) and larger than the critical density of spin
instability in neutron matter ($\varrho_c\approx
0.37\,\mbox{fm}^{-3}$). When some admixture of protons is added to
neutron matter, the last critical density is shifted to larger
densities and a spin polarized state with the oppositely directed
spins of neutrons and protons appears. Under increasing density of
nuclear matter, the neutron spin polarization continuously
increases till all neutron spins will be aligned in the same
direction. Protons, at first, become more polarized with density
and their spin polarization is opposite to the spin polarization
of neutrons. But, after reaching the maximum, spin polarization of
protons decreases and at some critical density spins of protons
change direction, so that the spin ordered phase with the
like-directed spins of neutrons and protons is formed. Then,
beyond the critical density, the spin polarization of protons is
continuing to increase until the totally polarized state with
parallel ordering of neutron and proton spins will be formed.
Thus, for SkI3 force  nuclear matter undergoes at some critical
density a phase transition from the state with antiparallel
ordering of neutron and proton spins to the state with  parallel
ordering of spins. With increasing isospin asymmetry, this
critical density increases as well. Note that there are no long
tails in the density profiles of the neutron spin polarization
parameter at large  asymmetries.

In Fig.~\ref{fig3}b  the total energies per nucleon of  spin
ordered and nonpolarized states are compared at zero temperature
for SkI3 force. In symmetric nuclear matter, FM spin ordering is
thermodynamically more preferable than AFM one. The energy gain of
the transition from the nonpolarized state to a spin polarized
state increases with isospin asymmetry at the given density,
analogously to SkI5 force. Hence,  a spin polarized state in
neutron matter occurs earlier in density than in symmetric nuclear
matter, as was clarified above. Note that for SkI3 force under
increasing density, initially, the state with antiparallel
ordering of neutron and proton spins appears without existence of
long tails in the density profiles of the neutron spin
polarization parameter at strong isospin asymmetry. This is in
contrast with SLy4 force, for which the antiparallel ordering at
$\alpha\lesssim 1$ is characterized by the appearance of such long
tails. Hence, the presence of long tails in the density profiles
of the neutron spin polarization parameter  doesn't relate to  the
antiparallel ordering of neutron and proton spins, but is
associated with  the decreasing dependence of the energy gain of
the phase transition to a spin polarized state as a function of
isospin asymmetry. In this case, the critical density of spin
instability in symmetric nuclear matter is less than the critical
density in neutron matter. If these critical densities are
substantially different, that is  the case for SLy4 force, then
the long tails in the density profiles of the neutron spin
polarization parameter appear.

In summary, we have considered  spin ordered phase transitions in
nuclear matter with SLy4, SkI3 and SkI5 Skyrme effective forces.
  It has
been shown    that asymmetric nuclear matter with SLy4 effective
interaction undergoes a phase transition to a state with the
oppositely directed spins of neutrons and protons. This phase
transition is characterized by the appearance of long tails in the
density profiles of the neutron spin polarization parameter near
the transition point at strong isospin asymmetry. This means, that
even small admixture of protons to neutron matter leads to the
considerable shift of the critical density of spin instability to
lower densities. The presence of such long tails is associated
with the decreasing dependence of the energy gain of transition to
a spin polarized state as a function of isospin asymmetry.

In  nuclear matter with SkI5 effective interaction   a spin
polarized state with the like-directed spins of neutrons and
protons is formed. For SkI3 force, nuclear matter  under
increasing density, at first, undergoes a phase transition to a
spin polarized state with the oppositely directed spins of
neutrons and protons. Under further increasing density, spins of
protons change their direction at some critical point and a phase
transition from the state with  antiparallel ordering to the state
with  parallel ordering of neutron and proton spins occurs. Since
the energy gain of the transition from the nonpolarized state to a
polarized one  is increased with isospin asymmetry, there are no
long tails in the density distribution of the neutron spin
polarization parameter for SkI3 and SkI5 forces. Note that the
obtained results may be of importance for the adequate description
of thermal and magnetic properties of neutron stars.

\end{document}